\documentclass[fleqn,usenatbib,twocolumn]{mnras}
\usepackage[utf8]{inputenc}
\usepackage{graphicx}
\usepackage{caption,subcaption}
\usepackage{amsmath}
\usepackage{siunitx}
\usepackage{comment}
\usepackage{mathtools}
\usepackage{lipsum}

\usepackage{listings}
\usepackage{url}

\def\prog[#1]#2{{\tt #1}}
\def\thetaE{\frac{R_E}{d_{LO}}}
\newcommand{\RE}{R_E} % Einstein radius in geometrical optics
\newcommand{\half}{{\textstyle\frac12}}
\newcommand{\muavr}{\langle\mu\rangle_{r_1}}
\newcommand{\muavth}{\langle\mu\rangle_{\theta_1}}

\title{Optical properties of the solar gravity lens} \author[S. Engeli
  \& P. Saha]{Sara Engeli\thanks{E-mail: sara.engeli@uzh.ch} and
  Prasenjit Saha\\Physik-Institut, University of Zurich,
  Winterthurerstrasse 190, 8057 Zurich, Switzerland}

\date{}

\begin{document}

\maketitle

\begin{abstract}
  It is well known that the solar gravitational field can be
  considered as a telescope with a prime focus at locations beyond
  550~au.  In this work we present a new derivation of the
  wave-optical properties of the system, by adapting the arrival-time
  formalism from gravitational lensing.  At the diffraction limit the
  angular resolution is similar to that of a notional telescope with
  the diameter of the Sun, and the maximum light amplification is
  $8\pi^4GM_\odot/(c^2\lambda)$, enough to detect a $1\,$W laser on
  Proxima Centauri~b pointed in the general direction of the Sun.
  Extended sources, however, would be blurred by the wings of the
  point spread function into the geometrical-optics regime of
  gravitational lensing.  Broad-band sources would have to further
  contend with the solar corona.  Imaging an exoplanet surface as
  advocated in the literature, without attempting to reach the
  diffraction limit, appears achievable.  For diffraction-limited
  imaging (sub-km scales from 100~pc) nearby neutron stars appear to
  be most plausible targets.
\end{abstract}

\begin{keywords}
\end{keywords}

\section{Introduction}

The phenomenon of light taking multiple paths through a gravitational
field is now a familiar one.  Small bright sources can produce two or
more discrete images when gravitationally lensed, as in the original
double quasar \cite[discovered by][]{1979Natur.279..381W}.  Extended
sources when similarly lensed produce arcs, which may blend nearly
into rings, for example the ``Cosmic Horseshoe'' \citep[discovered
  by][]{2007ApJ...671L...9B}.  Such systems are all in the regime of
geometrical optics; there is no optical interference between the
separate light paths.  The reason interference does not occur is that
the light paths differ in travel time by hours to years, which is much
longer than the coherence time of the light source.  Scenarios where
interference in gravitational lensing could occur have been studied
\citep[e.g.,][]{2020MNRAS.497.4956J,2021arXiv210909998R} but are not
observable yet.

There is, however, another way to see interference in gravitational
lensing, albeit a futuristic one. It involves using the Sun as the
lens, by sending an observer spacecraft to a distance $d_{OL}$ such
that the angular radius $R_\odot/d_{OL}$ of the Sun becomes smaller
than the deflection angle $4GM_\odot/(c^2 R_\odot)$ at the rim of the
Sun.  A light source precisely behind the Sun will then be lensed into
a diffracting ring, resulting in a real image with a point spread
function.  The required distance is $d_{OL}\geq\SI{550}{au}$ or about
three light days.  In the same year as the first gravitational lens
discovery and the first Saturn flyby \cite{1979Sci...205.1133E},
combining the fascinations of gravitational lensing and deep-space
missions, drew attention to both the great potential and the
formidable problems of a mission to the solar gravity focus.
Subsequently several other authors, notably \cite{2010AcAau..67..521M}
and recently \cite{2020arXiv200211871T} have advocated a solar gravity
lens mission.  The wave optics of the solar gravitational lens has
also been studied in several works
\citep{1976IJTP...15...45H,1986PhRvD..34.1708D,1999PThPS.133..137N,
  2013IJAA....3....1N, 2017PhRvD..95h4041T,2017PhRvD..96b4008T}.

In this paper we will re-derive the optical properties of the solar
gravity lens in a simple way, by adapting the Fermat-principle
formulation of gravitational lensing.  This approach most resembles
\cite{2013IJAA....3....1N}, whereas most other works proceed by
solving for a plane electromagnetic wave crossing a spherical
gravitational field.  We will then briefly discuss the expected photon
fluxes from different kinds of targets, and compare with the
foreground light from the solar corona.  The approach used here is
technically simpler, in that it involves a scalar quantity
(essentially the optical path length) rather than the electromagnetic
four-vector potential, but in doing so sacrifices information like
polarisation, which is encoded in the four-potential.

We will not attempt to address any spacecraft or instrument issues.
\cite{2020arXiv200211871T} is a good summary of these.  We will also
not include two important issues relating to the Sun.  One is possible
decoherence caused by the solar corona; \cite{2019PhRvD..99b4044T}
find that the effect is negligible at optical wavelengths.  The other
is perturbations due to the Sun's oblateness and higher multipoles;
this actually a significant effect
\citep[cf.][]{10.1093/ptep/pty119,2021PhRvD.104l4033T}, which we will
discuss briefly later, but for the present work we assume a spherical
Sun.

\section{Lensing time delays}

\begin{figure}
    \centering
    \includegraphics[width=\hsize]{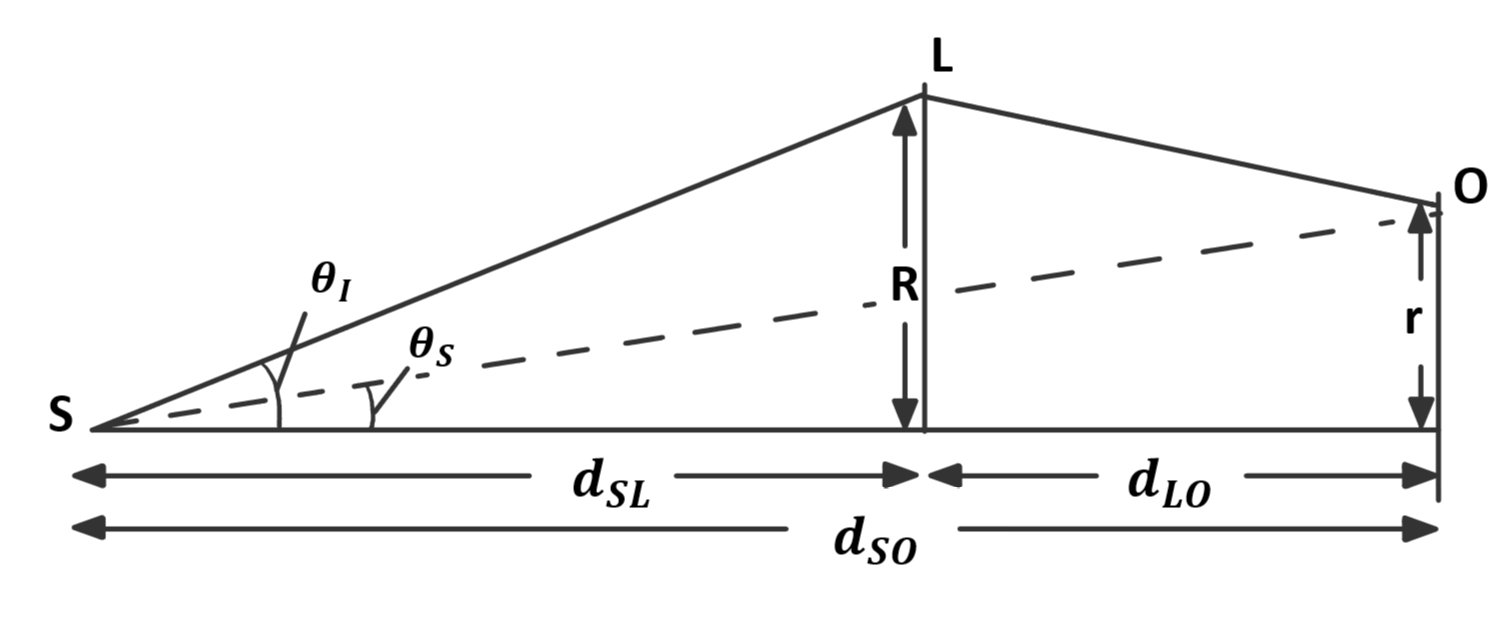}
    \caption{Distances and angles in gravitational lensing
      \citep[adapted from Fig.~1 in][]{1986ApJ...310..568B}.  The
      optical axis (horizonal line) runs from the source point through
      the centre of the lens to a reference point on the observer
      plane.  A photon leaves the source at angle $\theta_I$ from the
      optical axis, gets deflected and delayed at radius $R$ of the
      lens, and reaches the observer plane at radius $r$, which is the
      spot where an unlensed photon leaving the source at $\theta_S$
      would arrive.  The angles $\theta_I$ and $\theta_S$ are
      understood as two-dimensional.}
    \label{fig:fermatmod}
\end{figure}

Consider a possible path for a photon travelling from a source to a
point in the camera plane.  The path first goes in a straight line
to a point $(R,\Phi)$ in a plane through the sun parallel to the
camera plane; then it changes direction and takes a straight route to
the point $(r,\phi)$ in the camera plane.  Fig.~\ref{fig:fermatmod}
shows the geometry being considered, omitting $\Phi$ and $\phi$ for
simplicity.  It is the same as in the well-known formulation of
Fermat's principle in gravitational lensing by
\cite{1986ApJ...310..568B} except that source and observer have been
swapped.  From Eqs.~2.1--2.6 of that paper, the arrival time
\begin{equation}\label{fermatmodif}
  t(\theta_I, \theta_S) =  \frac{d_{SL}d_{SO}}{2cd_{LO}} (\theta_I-\theta_S)^2
    - 4\frac{GM_\odot}{c^3} \ln\theta_I
\end{equation}
follows, assuming the angles
\begin{equation}
  \theta_I = \frac{R}{d_{SL}} \qquad
  \theta_S = \frac{r}{d_{SO}}
\end{equation}
are small.  \cite{1986ApJ...310..568B} also include dependence on redshifts
in an expanding universe, which can be disregarded here.

Including the angles $\Phi$ and $\phi$ we have
\begin{equation}
\begin{aligned}
  t(r,\phi,R,\Phi) &= \frac{\eta r^2 + \eta^{-1} R^2
                            - 2rR \cos(\Phi - \phi)}{2c\,d_{LO}} \\
                 &- \frac{2R_S}c \, \ln\left(R/d_{SL}\right)
\end{aligned}
\end{equation}
where
\begin{equation}
R_S = \frac{2GM_\odot}{c^2}
\end{equation}
is the nominal Schwarzschild radius, and the distance ratio
$d_{SL}/d_{SO}$ is denoted by $\eta$.  For sources of interest,
$d_{SL}\gg d_{LO}$ and hence $\eta\approx 1$.  We can eliminate $\eta$
by redefining $R$ and $r$ slightly, simplifying the arrival time to
the following.
\begin{equation}
\begin{aligned}
  t(r,\phi,R,\Phi) &= \frac{r^2+R^2 - 2rR \cos(\Phi - \phi)}{2c\,d_{LO}} \\
                   &- \frac{2R_S}c \, \ln\left(R/d_{SL}\right)
\end{aligned}
\end{equation}
Since we have assumed a spherical Sun, the dependence on $\Phi$ and
$\phi$ is only through $\cos(\Phi-\phi)$.  Departures from a spherical
Sun will present a more complicated dependence \citep[cf.\ Eq.~119
  from][]{2021PhRvD.103f4076T}.

To get an expression for the amplitude in the observer plane we need
to sum up all virtual photon paths in the solar plane. This means we
integrate the photon paths over the solar plane:
\begin{equation}\label{integralpolar}
A(r,\phi) \propto \int e^{2\pi i \nu t(r,\phi,R,\Phi)} \, R\,dR\,d\Phi
\end{equation}
This is the Fresnel-Kirchhoff diffraction integral for our problem.

\begin{figure}
  \centering
  \includegraphics[width=\hsize]{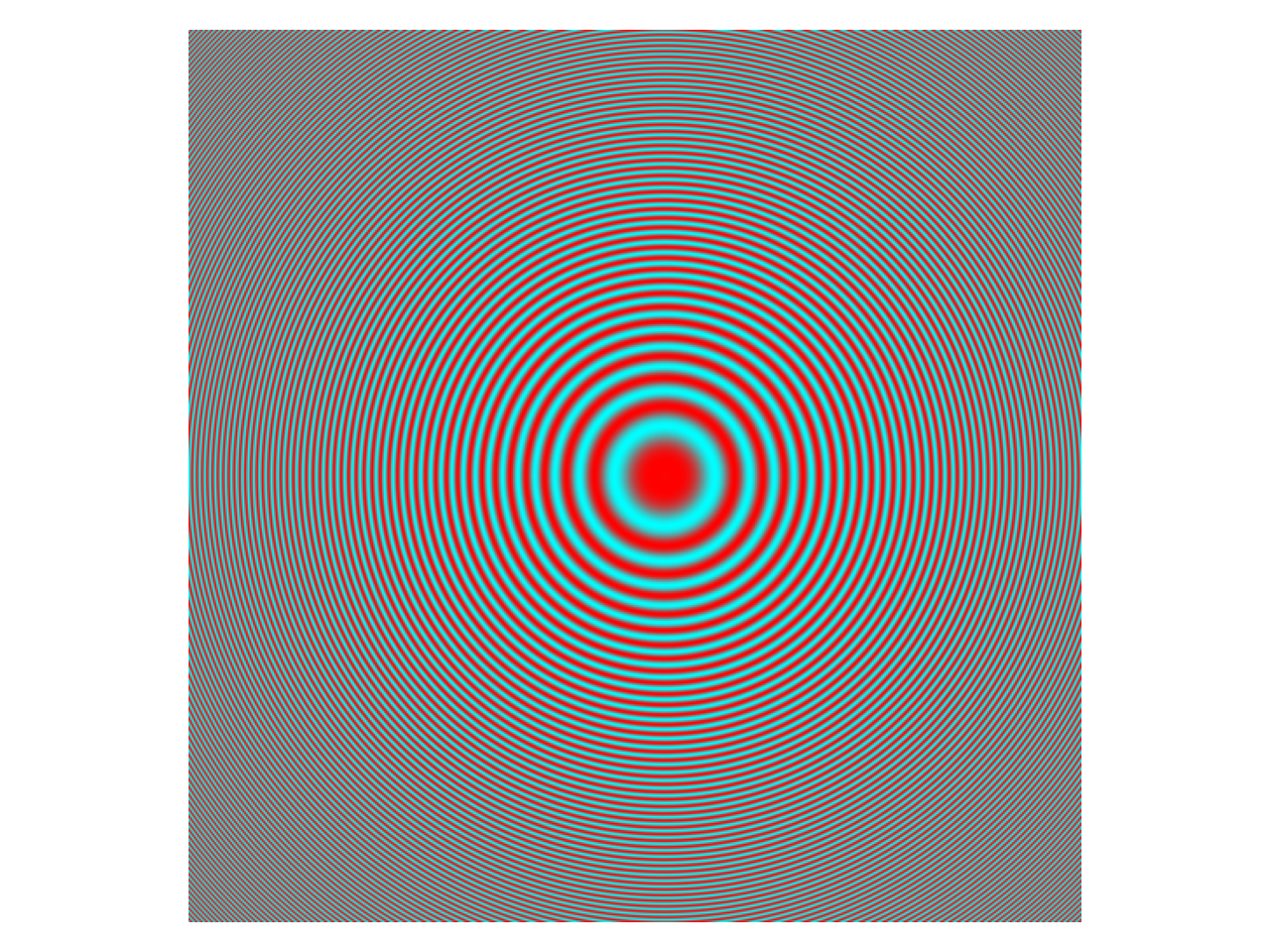}
  \includegraphics[width=\hsize]{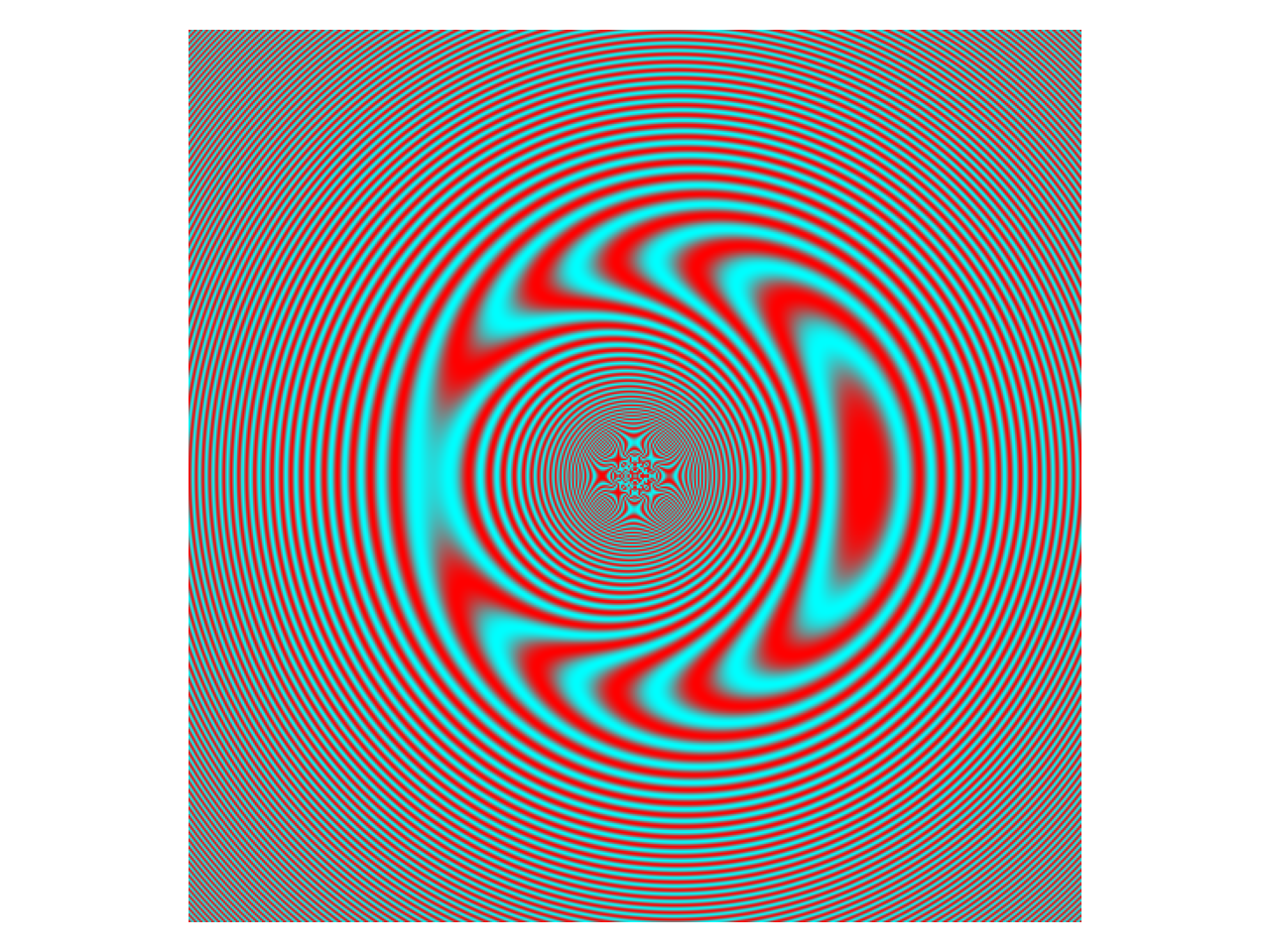}
  \caption{A representation of the phase in the observer's sky with no
    lens (upper panel) and with the solar gravity lens (lower panel).
    The quantity shown is $\cos(2\pi\nu t)$ from Eq.~\eqref{time},
    with red indicating $+1$, grey indicates 0, and cyan $-1$.  In an
    animated version of the figure, where the phase keeps increasing
    with time, the pattern appears to travel from higher-$t$ regions
    to lower-$t$ regions.  The observer is located at $r =
    \SI{2e8}{\metre}$ from the focal line, and $R$ in the figure
    ranges from 0 to \SI{1.4e9}{\metre}.  The Einstein radius $R_E$
    corresponds to the large ring in the lower panel.  The Fresnel
    scale $R_F$ corresponds to the largest fringes.  To make $R_F$
    noticeable, we have used a very low frequency $\nu=1\rm\,MHz$.  At
    optical wavelengths the Fresnel scale would be $\sim10^4$ times
    smaller.} \label{fig:timearriv}
\end{figure}

To simplify the following derivation, it is useful to introduce
\begin{equation}
\begin{aligned}
R_E &= \sqrt{2R_S\,d_{LO}} \\
R_F &= \sqrt{\lambda d_{LO}}
\end{aligned}
\end{equation}
which are the Einstein radius and the Fresnel scale respectively.  We
now rewrite the phase as
\begin{equation}\label{time}
\begin{aligned}
  2\pi\nu t(r,\phi,R,\Phi) = \frac{2\pi}{R_F^2}
  \Big( &\half(r^2 + R^2) - rR \cos(\Phi - \phi) \\
        &- R_E^2 \, \ln(R/d_{SL}) \Big)
\end{aligned}
\end{equation}
Fig.~\ref{fig:timearriv} illustrates the arrival time \eqref{time},
without and with the last (lensing) term.  The quantity shown is
$\cos(2\pi\nu t)$ for $t(R,\Phi)$ at a fixed $r,\phi$ and a notional
$\nu$ of $1\rm\,MHz$.  A grey blur in the figure indicates $t$ varying
quickly with $R,\phi$.  Extended red or cyan regions in the figure
indicates where $t$ is stationary or nearly stationary.  In the
absence of a lensing mass, the only stationary point of $t$ is a
minimum, as evident in the upper panel.  In the lower panel, a minimum
and a saddle point are apparent.  In geometrical optics, there are
images (virtual images) at such stationary points of $t$.  In wave
optics, however, we need to integrate the complex amplitude over $R$
and $\Phi$.

\section{The point spread function}

The phase in the integrand in \eqref{integralpolar} can be split up in
a part of angular dependence and a part of radial dependence.  We
discard constants and dependencies on $r$ only, which contribute only
constant phase.  This leads to a split up integral for the
amplitude:
\begin{equation}\label{integsplit}
A(r,\phi) \propto
  \int e^{i\psi(R)} R\,dR \,
  \int e^{-2\pi i (rR/R_F^2) \cos(\Phi - \phi)} d\Phi
\end{equation}
where
\begin{equation}\label{phase}
\psi(R) = \frac{2\pi}{R_F^2} \Big(\half R^2 - R_E^2 \ln(R/d_{OL}) \Big)
\end{equation}
Using the well-known integral for Bessel functions
\begin{equation}\label{besselfunction}
J_0(u) = \frac{1}{2\pi} \int_{-\pi}^{\pi} e^{i u \cos(\Phi)} d\Phi
\end{equation}
the double integral in \eqref{integsplit} can be simplified to a
single integral:
\begin{equation}\label{integbessel}
  A(r) \propto \int e^{i\psi(R)} J_0\left(\frac{2\pi rR}{R_F^2}\right)
  \, R\, dR
\end{equation}
Integration over $\Phi$ also removes the $\phi$ dependence, because
only the difference $\Phi-\phi$ appears in the integrand in
Eq.~\eqref{integsplit}.

The phase \eqref{phase} has a minimum at $R=R_E$.  Since only the
region of minimal phase change contributes to the image, we can
approximate the phase function around its minimum in order to get a
simpler expression.  With a Taylor expansion the phase function
\eqref{phase} can be approximated around $\RE$:
\begin{equation}\label{taylor}
\begin{aligned}
\psi(R) &= \psi(R_E) + \psi'(R_E) (R-R_E) \\
          &+ {\textstyle\frac12} \psi''(R_E)(R-R_E)^2 + O((R-R_E)^3)
\end{aligned}
\end{equation}
In \eqref{taylor} we note that $\psi'(R_E)=0$.  Further, terms not depending
on $r$ or $R$ can be ignored, since $\psi$ is a phase. In our case $O((R-R_E)^3)$ until $R = R_E \pm 20 R_F$ is still about $10^4$ times smaller then the previous terms in the Taylor approximation.  Since the region of $R$ we are integrating over is always within $R = R_E \pm 20 R_F$ we can ignore $O((R-R_E)^3)$ and terms of higher order. The phase can then be approximated as
\begin{equation}\label{phaseapprox}
  \psi(R) \approx 2\pi \frac{(R-R_E)^2}{R_F^2}
\end{equation}
We can take the terms that do not depend on $R$ out of the integral
\eqref{integbessel} and get:
\begin{equation}\label{intensity_integralsimp}
  A(r) \propto \int_{-\infty}^\infty e^{2\pi i\left(\frac{R-R_E}{R_F}\right)^2}
  J_0\left(\frac{2\pi r R}{R_F^2}\right) \, R\,dR
\end{equation}

To find the light amplification $\mu$ we have to calculate
\begin{equation}\label{mu}
    \mu = \left|\frac{A(r)}{A_0(r)}\right|^2
\end{equation}
where $A_0(r)$ is the diffraction integral for the case of no lens.
Here we cannot simply put $R=R_E$ in the expression
\eqref{intensity_integralsimp} for $A(r)$, because the
Taylor-approximated phase \eqref{phaseapprox} is not valid for
$R_E=0$.  We have to go back to the earlier expression
\eqref{integbessel} and put $R_E=0$ there.  This gives
\begin{equation}
  A_0(r) \propto \int_{-\infty}^\infty e^{\pi i\left(\frac{R}{R_F}\right)^2}
  J_0\left(\frac{2\pi r R}{R_F^2}\right) \, R\,dR
\end{equation}
We thus have
\begin{equation}
  \frac{A(r)}{A_0(r)} =
  \frac{\int_0^\infty e^{2\pi i\left(\frac{R-R_E}{R_F}\right)^2}
         J_0\left(\frac{2\pi rR}{R_F^2}\right) \, R\,dR}
       {\int_0^\infty e^{\pi i\left(\frac{R}{R_F}\right)^2}
         J_0\left(\frac{2\pi r R}{R_F^2}\right) \, R\,dR}
\end{equation}
In the numerator the integrand contributes significantly only near
$R=R_E$, so we can take any slow $R$ dependence outside the integral.
This lets us simplify to the following.
\begin{equation}
\begin{aligned}
  \frac{A(r)}{A_0(r)} ={} &\frac{2R_E}{R_F} \,
  J_0\!\left(\frac{2\pi rR_E}{R_F^2}\right) \times \\
  &\frac{\int_0^\infty e^{2\pi i u^2} \,du}
        {\int_0^\infty e^{\pi i u^2}
          J_0\!\left(\frac{2\pi r u}{R_F^2}\right) \, u\,du}
\end{aligned}
\end{equation}
Substituting the standard integrals
\begin{equation}
\begin{aligned}
  & \int_0^\infty e^{2\pi i u^2} \,du = (1+i)/4 \\
  & \int_0^\infty e^{\pi i u^2} J_0(ku) \, u\,du =
    i e^{ik^2/(4\pi)} / (2\pi)
\end{aligned}
\end{equation}
and simplifying gives the normalised light amplification
\begin{equation}\label{musimp}
  \mu = 4\pi^2 \frac{R_S}\lambda
  J_0^2\left(\frac{2\pi r}{\lambda}\thetaE\right)
\end{equation} 
which, since it is a function of radius $r$ on the observer plane, can
be considered a point spread function.

We can compare our result \eqref{musimp} with the expression in Eq.~135 of
\cite{2017PhRvD..96b4008T}.  Substituting $\rho \rightarrow r$, $r_g
\rightarrow R_S$ and $z \rightarrow d_{LO}$ in their expression gives
\begin{equation}\label{mu_ref}
  \mu = \frac{4\pi^2 R_S/\lambda}{1-e^{-4\pi^2 R_S/\lambda}}
  J_0^2\left(\frac{2\pi r}{\lambda}\thetaE\right)
\end{equation} 
The exponential function in \eqref{mu_ref} for typical wavelengths
$\lambda$ in the visible spectrum is almost 0. So \eqref{mu_ref}
simplifies to the same as \eqref{musimp}.

\section{The diffraction limit vs geometrical optics}

If we look at a source plane instead of a point source, it is more useful to know the average amplification $\langle \mu \rangle$ of the whole observation area. For a circular source plane we can change the integral to polar coordinates and we find the average amplification as a function of the maximal radius $r_1$ in the observation plane.

\begin{equation}\label{muav}
 \muavr = \frac{2\pi \int_0^{r_1} \mu(r) \,r \,dr}{\pi r_1^2}
\end{equation}
Before inserting \eqref{musimp} in \eqref{muav}, the Bessel function in the former can be substituted by an approximation that is easier to integrate. For large argument $J_0$ can be approximated \citep[see e.g.,][Eq.~11]{doi:10.1080/00029890.1963.11992149} as
\begin{equation}\label{besselasymp}
 J_0(x) \approx \sqrt{\frac{2}{\pi x}} \cos{\left(x - \frac{\pi}{4}\right)}
\end{equation}
We then insert \eqref{musimp} and \eqref{besselasymp} in \eqref{muav} and 
simplify to get
\begin{equation}\label{muav3}
 \muavr = \frac{4 R_E}{r_1^2} 
 \int_0^{r_1}  \cos^2{\left(\frac{2\pi r}{\lambda}\thetaE - \frac{\pi}{4}\right)} \,dr 
\end{equation}
The integral in \eqref{muav3} can now be calculated analytically and gives us the resulting formula for the average amplification:
\begin{equation}\label{muav4}
  \muavr = \frac{2 R_E}{r_1} +
  \frac{\lambda d_{LO}}{2\pi r_1^2}
  \left(1-\cos\left(\frac{4\pi r_1}{\lambda}\thetaE\right) \right)
\end{equation}
As we see, the amplification of extended sources tends to become independent
of wavelength.

\begin{figure}
  \centering
  \includegraphics[width=\hsize]{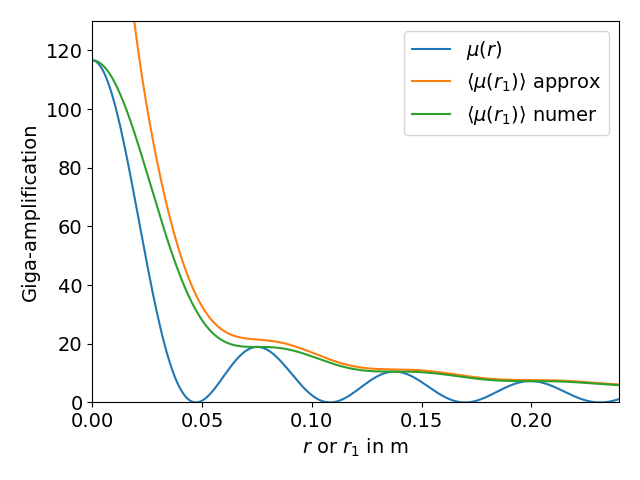}
  \caption{The point spread function and its enclosed averages.  The
    plotted quantities are $\mu(r)$ from Eq.~\eqref{musimp}, and its
    enclosed average $\langle\mu(r_1)\rangle$ as computed numerically
    from Eq.~\eqref{muav} and as estimated in Eq.~\eqref{muav4}.
    \label{fig:amplif}}
\end{figure}

Let us now change to angular terms.  Let
\begin{equation}
\theta_1 = \frac{r_1}{d_{LO}} = \frac{R_1}{d_{SL}}
\end{equation}
be the angular radius of the source, and
\begin{equation}
\theta_E = \frac{R_E}{d_{LO}}
\end{equation}
the angular Einstein radius.  In terms of these, the diffraction limit
is given by
\begin{equation}\label{difflimitcond}
J_0\left(\frac{2\pi R_E}\lambda \theta_\lambda\right) = 0
\end{equation}
\begin{equation}
\theta_\lambda = 0.77 \frac\lambda{2R_E}
\end{equation}
Recall that an ordinary telescope has $J_1$ instead of $J_0$ in the
condition \eqref{difflimitcond}.

For sources much larger than the diffraction limit but still small
\begin{equation}
\theta_\lambda \ll \theta_1 \ll \theta_E
\end{equation}
the mean amplification is
\begin{equation}\label{muav5}
\muavth = \frac{2\theta_E}{\theta_1} 
\end{equation}
Now, for gravitational lensing by a star in geometrical-optics, there
is a well-known expression for amplification that goes back to
\cite{1936Sci....84..506E}
\begin{equation}
  \frac1u \frac{1+u^2/2}{\sqrt{1+u^2/4}} \quad \hbox{where} \quad
  u = \frac{\theta_1}{\theta_E}
\end{equation}
Averaging over a disc of radius $\theta_1$ (assuming
$\theta_1\ll\theta_E$) gives the same as Eq.~\eqref{muav5}.

Thus, as we might have expected, geometrical optics applies for
sources much larger than the diffraction limit.

\section{Foreground and background}

If the Sun were dark, no further mirrors or lenses would be needed, a
detector to gather light would be sufficient.  Covering up the Sun,
however, and letting through the light in the Einstein ring,
necessitates a spacecraft telescope with some kind of coronograph.

The Sun from $\SI{600}{\rm au}$ has a spectral photon flux
\begin{equation}
\Phi_\lambda({\rm Sun}) \approx \SI{1e13}
{{\rm photons\;}\second^{-1}\metre^{-2}\nano\metre^{-1}}
\end{equation}
at $\SI{1}{\micro\metre}$.  This is comparable to the brightness of
the Moon as seen from the Earth, but over a very small area of
$\sim\SI{10}{\rm arcsec^2}$.  Thus the optical setup would be very
different from an ordinary coronograph, and more like the occulting
masks developed for imaging extrasolar planets \citep[see
  e.g.,][]{2020PASP..132a5002B}.  We emphasise, however, that the idea
is not to image the Einstein ring, but to let the amplitude through a
narrow ring around $R_E$ interfere.

The width of the ring of light could be reduced to the diffraction
limit of the observing telescope.  Mission concepts envisage a 1~m
telescope, which at optical wavelengths implies a ring of thickness
$\approx 0.1''$.  With $R_E\approx2''$ the area of the ring would be
about $\SI{1}{\rm arcsec^2}$, or about a tenth of the area of the
solar disc.  Coming through this ring would be light from the solar
corona.  Close to the Sun, the surface brightness of the corona is a
few times $10^{-6}$ that of the solar disc
\citep{1996ApJ...466..512N}.  Thus suggests
\begin{equation}
\Phi_\lambda({\rm solar\ corona}) \sim 10^6 \si{{\rm
    photons\;}\second^{-1}\metre^{-2}\nano\metre^{-1}}
\label{Phi-corona}
\end{equation}
which is comparable to a bright star.  The corona brightness itself
could be subtracted out, but the shot noise from it will remain as a
noise source, as will any intrinsic variation in the corona
brightness.

The light ring will also let in some light from the night sky.  That
light will be lensed, but because lensing preserves surface
brightness, the resulting photon flux will be the same as the unlensed
night sky.  The night sky brightness is $\approx\SI{3e12}{{\rm
    photons\;}\metre^{-2}\second^{-1}\steradian^{-1}}$ in a broad band
\citep[e.g., Table~1 from][]{2002NIMPA.481..229P}.  Since $\SI{1}{\rm
  arcsec^2}\simeq\SI{2.5e-11}{\rm sr}$ our assumed light ring will let
in $\sim \SI{10}{{\rm photons\;}\second^{-1}\metre^{-2}}$ even in a
broad band.

Thus the night-sky background is negligible, while the solar corona
would be the principal limitation.

\begin{figure}
  \centering
  \includegraphics[width=\hsize]{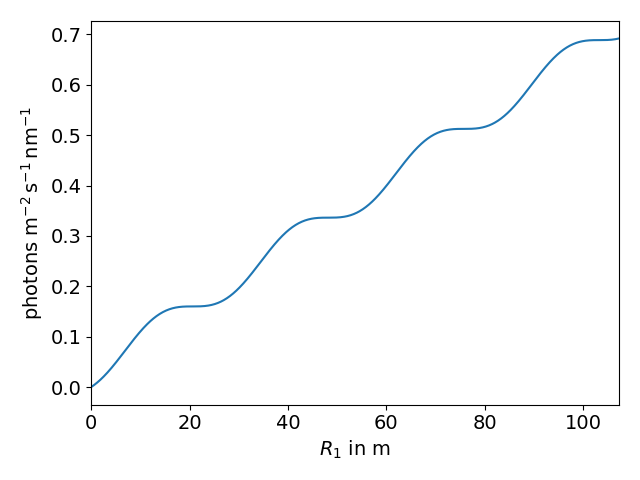}
  \caption{Observable photon flux at $\SI{1}{\nano\metre}$ from a
    uniform-brightness disc of radius $R_1$ at $\SI{1.3}{\rm pc}$
    emitting $\SI{1}{\watt\metre^{-2}\nano\metre^{-1}}$.}
    \label{fig:phflux}
\end{figure}

\section{Plausible targets}

As potential targets, exoplanets in habitable zones immediately come
to mind, especially well known being Proxima Centauri~b at $1.3\,$pc
\citep{Anglada_Escud__2016}, Teegarden~b at $3.8\,$pc
\citep{Zechmeister_2019}, Trappist-1d at $12.1\,$pc
\citep{Gillon_2016} and TOI~700~d at $31\,$pc \citep{Gilbert_2020}.
The last of these is a transiting system, and hence its orbital
inclination is measured and its exact position can be predicted.  The
other three are inferred from radial-velocity perturbations of the
host star, and their orbital inclinations are unknown.  As a result,
the location of the $r=0$ point on the observer plane has a large
uncertainty.  This problem could, however, be solved by preliminary
direct imaging of the planet at the single-pixel level, which can be
expected long before a 550~au mission becomes feasible.

The surface brightness of these exoplanets is not known, but
$\sim\SI{1}{\watt\metre^{-2}\nano\metre^{-1}}$ is a convenient round
figure to assume for the spectral flux. The solar spectral irradiance
in the visible range is close to this value
\citep[e.g.,][]{2004SoEn...76..423G}.  Converting to a photon rate and
multiplying by the brightness factor (\ref{muav4}) gives the
observable spectral photon flux, say $\Phi_\lambda$ in $\si{{\rm
    photons\;}\second^{-1}\metre^{-2}\nano\metre^{-1}}$.
Fig.~\ref{fig:phflux} shows the result at $\SI{1}{\nano\metre}$ for
$d_{SL}=1.3\rm\,pc$ and $d_{LO}=600\rm\,au$.  It is evident that even
small bodies would yield non-zero photons.

The extended wings of the point spread function, and furthermore the
bright foreground from the solar corona, would considerably degrade
the achievable resolution of exoplanets, compared to the incredible
$\SI{20}{\metre}$-scale on Proxima Centauri~b suggested by
Figs.~\ref{fig:amplif} and \ref{fig:phflux}.
\cite{2021PhRvD.103l4038T} carry out image deconvolution on a
simulated Earth at the distance of Proxima, to a resolution of roughly
$\SI{100}{\kilo\metre}$, and \cite{2022MNRAS.tmp.2011T} consider
deconvolution from 1200~au, where the solar corona is much fainter.
Extrapolating from Fig.~\ref{fig:phflux} we can estimate $\sim
10^3\si{{\rm photons\;}\second^{-1}\metre^{-2}\nano\metre^{-1}}$ from
a 100~km pixel on Proxima Centauri~b.  Comparing with the foreground
\eqref{Phi-corona} we see that the exoplanet will be orders of
magnitude fainter than the solar corona, but not that much fainter
than the \textit{noise} in the solar corona.  These estimates, though
of course only very rough, indicate that the solar corona would not
prevent the imaging of exoplanets.

For imaging at the diffraction limit, the best prospect would be an
isolated neutron star.  The nearest of these \citep[see
  e.g.,][]{2013ffep.confE...6H} is RX~J1856.5$-$3754, about $130\,$pc
away.  The distance is 100 times further than Proxima, but observing
at a much shorter wavelength of say $\SI{100}{\nano\metre}$ would be
desirable, giving a resolution of $\approx\SI{200}{\metre}$ at the
diffraction limit.

Laser lines are interesting, because they could be observed in a very
narrow band (say $\SI{0.01}{\nano\metre}$) thereby greatly reducing
the foreground.  Natural laser lines are known \citep[in
  $\eta\,$Carinae, see][]{2007NewAR..51..443J} but why not artificial
laser lines sent by interstellar friends?  Various SETI scenarios have
been discussed in \cite{2018AcAau.142...64H}.  Here we add one more.

Consider the first plateau in Fig.~\ref{fig:phflux}, which indicates
the diffraction limit.  This corresponds to a circular area of radius
$R_1=\SI{20}{\metre}$ or about $\SI{1000}{\metre^2}$ on Proxima~b.
Over a bandwidth of $1\rm\,nm$, this area at its assumed brightness
emits $1\rm\,kW$ of light and gives the observer $\sim \SI{0.15}{{\rm
    photons}\;\second^{-1}\metre^{-2}}$.  Now imagine a $1\rm\,W$
laser with a milliradian dispersion, located anywhere inside this
area, and aimed within a milliradian of the Sun.  With the light
within $10^{-6}\pi\rm\,sr$ rather than $2\pi\rm\,sr$ as with ordinary
light, the laser would be equivalent to $2\rm\,MW$ of ordinary light
emitted, and $\sim \SI{300}{{\rm photons}\;\second^{-1}\metre^{-2}}$
at the observer.  Absent the foreground from the solar corona, this
level of photon flux would be easily detectable.  Through the solar
corona, the laser would have to shine for some time, perhaps as short
as $\SI{10}{\second}$, to be detectable.  There are complications
arising from the small but non-zero asphericity of the Sun, which will
spread the light out over a caustic pattern
\citep{,2021PhRvD.104b4019T,2021PhRvD.104d4032T}, which we have not
investigated, but it appears plausible that with the solar gravity
lens, we could detect a laser pointer on Proxima~Centauri~b aimed
towards the Sun.  Provided of course, that we knew precisely where to
look.

\section*{Acknowledgements}
We thank G.~F.~Lewis, V.~Toth, P.~Tuthill, L.~L.~R.~Williams,
O.~Wucknitz, and the referee for comments.

\section*{Data Availability}
The code to generate the simulated data and figures are included in
the supplementary material.

\bibliographystyle{mnras}
\bibliography{main.bib}

\end{document}